\documentclass[amsmath,amssymb,prl,twocolumn,showpacs]{revtex4}
\usepackage[dvips]{graphicx}
\usepackage{bm}% bold math

\begin{document}
\title{Fluctuation Theorem for a Small Engine and Magnetization Switching by Spin Torque}
\author{Yasuhiro Utsumi}
\address{Department of Physics Engineering, Faculty of Engineering, Mie University, Tsu, Mie, 514-8507, Japan}

\author{Tomohiro Taniguchi}
\address{National Institute of Advanced Industrial Science and Technology (AIST), Spintronics Research Center, 1-1-1 Umezono, Tsukuba 305-8568, Japan}

\begin{abstract}
We consider a reversal of the magnetic moment of a nano-magnet by the fluctuating spin-torque induced by a non-equilibrium current of electron spins. 
This is an example of the problem of the escape of a particle from a metastable state subjected to a fluctuating non-conservative force. 
The spin-torque is the non-conservative force and its fluctuations are beyond the description of the fluctuation-dissipation theorem. 
We estimate the joint probability distribution of work done by the spin torque and the Joule heat generated by the current, which satisfies the fluctuation theorem for a small engine. 
We predict a threshold voltage above which the spin-torque shot noise induces probabilistic switching events and below which such events are blocked. 
We adopt the theory of the full-counting statistics under the adiabatic pumping of spin angular momentum. 
This enables us to account for the backaction effect, which is crucial to maintain consistency with the fluctuation theorem. 
\end{abstract}

\date{\today}

%05.40.-a	Fluctuation phenomena, random processes, noise, and Brownian motion (for fluctuations in superconductivity, see 74.40.-n; for statistical theory and fluctuations in nuclear reactions, see 24.60.-k; for fluctuations in plasma, see 52.25.Gj; for nonlinear dynamics and chaos, see 05.45.-a)

%72.70.+m	Noise processes and phenomena
%75.78.-n	Magnetization dynamics
%75.60.Jk	Magnetization reversal mechanisms

%85.75.-d	Magnetoelectronics; spintronics: devices exploiting spin polarized transport or integrated magnetic fields
%72.25.-b	Spin polarized transport (for spin polarized transport devices, see 85.75.-d)
%05.40.Jc	Brownian motion

\maketitle

\newcommand{\mat}[1]{\mbox{\boldmath$#1$}}

%\section{Introduction}

\begin{figure}[ht]
\begin{center}
\includegraphics[width=1.0 \columnwidth]{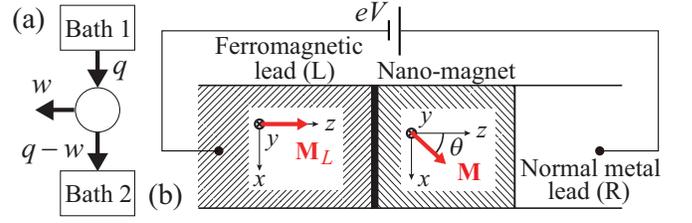}
\caption{
(a) Schematic picture of a small engine. 
The input heat $q$ and the output work $w$ fluctuate. 
(b)
A nano-magnet coupled to the left ferromagnetic lead and the right normal metal lead. 
The directions of magnetic moments of the ferromagnetic lead and the nano-magnet are 
${\mathbf e}_z=(0,0,1)$ 
and 
${\mathbf m}=(\sin \theta \cos \phi, \sin \theta \sin \phi, \cos \theta)$. 
}
\label{fig:fifin}
\end{center}
\end{figure}

The thermodynamics of small systems, the stochastic thermodynamics~\cite{Seifert}, is of growing importance in nano-science. 
The key ingredient is the fluctuation theorem (FT)~\cite{Seifert,EHM,CHT}, which has been applied to the solid state physics recently and extends the fluctuation-dissipation theorem as well as the Onsager relations far from equilibrium (see e.g. Refs.~\onlinecite{EHM,CHT,Tobiska,FB,SU,Andrieux,UI,Lopez}). 
Recent studies suggest that the FT is also useful to analyze small engines~\cite{Sinitsyn,Campisi,Verley}. 
In a small engine, during a short time step $\Delta t$ at finite temperature $T$, the input heat $q$ and the output work $w$ fluctuate and can take positive and negative values [Fig. \ref{fig:fifin} (a)]. 
The FT ensures that the joint probability distribution satisfies 
%----------------------------------------------------------
\begin{eqnarray}
P_{R, \Delta t}(-q,-w)
=
P_{\Delta t}(q,w)
e^{-\beta (q+w)}, 
\;
\beta=(k_{\rm B}T)^{-1},
\label{ftsmen}
\end{eqnarray}
%----------------------------------------------------------
where the subscript $R$ indicates that the external driving is reversed. 
From Jensen's inequality, this equation reproduces the Carnot theorem,  
$
\langle w \rangle/\langle q \rangle
\leq 1
$. 
The FT~(\ref{ftsmen}) is applicable even when a cycle is not defined. 
The work can be attributed to a non-conservative force originating from a heat flow between two baths [Fig. \ref{fig:fifin} (a)]. 
Let us couple the small engine to a small system. 
The energy variation of the small system is equal to the fluctuating work:  
%----------------------------------------------------------
\begin{eqnarray}
\Delta E =w
\, . 
\label{laneqn}
\end{eqnarray}
%----------------------------------------------------------
We expect that Eqs.~(\ref{ftsmen}) and (\ref{laneqn}) are applicable to a wide spectrum of mesoscopic systems driven by non-conservative forces. 

In the present paper, we apply this idea to the problem of the escape of a particle from a metastable state~\cite{HTB} subjected to a fluctuating non-conservative force. 
We consider the following specific setup:
a nano-magnet connected to a left ferromagnetic lead (source) and a right normal metal lead (drain)  [Fig. \ref{fig:fifin} (b)]. 
The magnetization vector of the bulk left ferromagnetic lead ${\mathbf M}_L$ is fixed. 
Let us assume that the magnetization of the nano-magnet ${\mathbf M}$ is anti-parallel to ${\mathbf M}_L$. 
By applying a source-drain bias voltage $V$, spin polarized electrons are injected from the ferromagnetic lead, which exert a torque on the nano-magnet~\cite{Slonczewski}. 
When the magnetic moment ${\mathbf M} {\mathcal V}$ (${\mathcal V}$ is the volume of the nano-magnet) is small, above a critical voltage $V^*$, ${\mathbf M}$ is reversed and aligns parallel to ${\mathbf M}_L$.  
The spin-torque is generated by the non-equilibrium current and thus the non-conservative force. 
It performs the work $w$ on the small system (the nano-magnet) and is accompanied by the Joule heat $q$. 
Since the spin angular momentum exchanged between electrons and the nano-magnet is discretized by $\hbar$, the spin-torque fluctuates and even under the critical voltage $V^*$, it can switch the magnetic moment probabilistically. 
The exponent $\Delta$ of the switching probability 
%----------------------------------------------------------
\begin{eqnarray}
P_\tau \sim e^{-\Delta}
\, , 
\label{swipro}
\end{eqnarray}
%----------------------------------------------------------
is well studied for equilibrium thermal fluctuations, which are Gaussian-distributed (see e.g. Refs.~\onlinecite{Apalkov,TI,Pinna,TUMGI} and references therein). 
However, this is not the case for the non-equilibrium fluctuations. 
In current experiments~\cite{Suzuki}, an MgO-insulating tunnel barrier is sandwiched between the nano-magnet and the ferromagnetic lead, which generates a Poisson-distributed shot-noise out of equilibrium~\cite{Blanter}. 
Previous studies analyzing the non-equilibrium spin-torque shot noise~\cite{Foros,Nunez,Chudnovskiy} limited themselves to the Gaussian fluctuations. 
The non-Gaussian fluctuations are beyond the description of the fluctuation-dissipation theorem and, to our knowledge, have not been reliably described.  

In the present paper, we determine the distribution of non-Gaussian fluctuations by using the full-counting statistics~\cite{Levitov} under the adiabatic pumping~\cite{Andreev,Kamenevbook}, which gives the joint probability distribution consistent with the FT for a small engine (\ref{ftsmen}).  
We evaluate the switching exponent $\Delta$ and predict another threshold voltage $V_{\rm th} $ under which the probabilistic switching is completely blocked. 
This is a result of the backaction, i.e., the adiabatic pumping of the spin angular momentum~\cite{Brataas}, as a consequence of the FT.

{\it Langevin equation in the energy coordinate --}
We take the $z$-axis parallel to the direction of the left magnetization, 
${\mathbf e}_z=(0,0,1)={\mathbf M}_L/|{\mathbf M}_L|$, 
which is fixed [Fig. \ref{fig:fifin} (b)]. 
We assume the uniaxial anisotropy of the nano-magnet in the $z$-direction. 
The anisotropic energy is, 
%----------------------------------------------------------
\begin{eqnarray}
E
=
-\frac{M H_K {\mathcal V} ({\mathbf e}_z \cdot {\mathbf m})^2}{2}
=
-\frac{M H_K {\mathcal V} \cos^2 \theta}{2}
\, , 
\label{energy}
\end{eqnarray}
%----------------------------------------------------------
where $M=|{\mathbf M}|$ is the saturation magnetization and ${\mathbf m}= {\mathbf M}/M$. 
In the spherical coordinates,  it is expressed as 
${\mathbf m} = (\sin \theta \cos \phi, \sin \theta \sin \phi, \cos \theta)$. 
The anisotropic magnetic field is typically $H_K>0$, and thus the magnetic moment tends to align with  ${\mathbf m}={\mathbf e}_z$ or ${\mathbf m}=-{\mathbf e}_z$. 
These 2 states are separated by the energy barrier $M H_K {\mathcal V}/2$. 
Because of this bistability, the setup is applicable to a memory device~\cite{Suzuki}. 

The dynamics of the nano-magnet is described by the stochastic Landau-Lifshitz-Gilbert equation, 
%----------------------------------------------------------
\begin{eqnarray}
\dot{{\mathbf m}}
=
-\gamma 
{\mathbf m}
\times
(
{\mathbf H}_{\rm eff}
+
{\mathbf h}
)
+
\alpha \, 
{\mathbf m}
\times
\dot{{\mathbf m}}
-
\gamma \, {\mathbf I}_S /(M {\mathcal V})
\, , 
\label{LLG}
\end{eqnarray}
%----------------------------------------------------------
where 
$\gamma = 2 \mu_{\rm B}/\hbar$ is the gyromagnetic ratio and $\mu_{\rm B}$ is the Bohr magneton. 
The effective magnetic field is 
%----------------------------------------------------------
$
{\mathbf H}_{\rm eff}
=
-{\mathcal V}^{-1}
{\partial E}/\partial {\mathbf M}
=
H_K \cos \theta {\mathbf e}_z
$, 
%----------------------------------------------------------
and ${\mathbf h}$ is its fluctuation induced by thermally excited magnons. 
It is a Gaussian white noise, i.e., $\langle {h}_j(t) \rangle = 0$ ($j=x,y,z$),
and the correlation is instantaneous and isotropic: 
$\langle {h}_j(t) {h}_k(t') \rangle 
=
2 \alpha k_{\rm B}T \, 
\delta_{jk} \, \delta(t-t')
/(\gamma M {\mathcal V})$. 
The Gilbert damping constant $\alpha$ also appears in the second term of the lhs of Eq.~(\ref{LLG}), indicating the relaxation to ${\mathbf m} = {\mathbf e}_z$ or ${\mathbf m} = - {\mathbf e}_z$. 
The spin-torque 
${\mathbf I}_S
=
{\mathcal I}
{\mathbf m}
\times
({\mathbf e}_z
\times
{\mathbf m}
)$ 
aligns ${\mathbf M}$ parallel to ${\mathbf M}_L$~\cite{Slonczewski}. 

%----------------------------------------------------------
%----------------------------------------------------------

Since typically the damping and the spin torque are weak, the variation of the energy after a single precession is small~\cite{Apalkov,Pinna,TUMGI}. 
Therefore, the magnetic moment precesses along the $z$ axis with the frequency 
$\dot{\phi}=\gamma H_K \cos \theta \equiv \Omega$ along a constant energy trajectory given by Eq.~(\ref{energy}). 
In the following, we will concentrate on the negative branch, $\Omega=-\sqrt{-2 \gamma^2 H_K E/(M {\mathcal V})}$, i.e., $-1 \leq m_z \leq 0$. 
It is convenient to consider the time derivative of the energy (\ref{energy}) averaged over a single precession:
$\overline{ \dot{E}(t) }
=
\Omega
\int_t^{t+2 \pi/\Omega} dt'
\dot{E}(t')
/(2 \pi)$. 
In the first order in $\alpha$ and ${\mathcal I}$, we obtain Eq.~(\ref{laneqn}) for our system:
%----------------------------------------------------------
\begin{eqnarray}
\overline{ 
\dot{E} 
}
=
\overline{ 
\dot{ {\mathbf M} } \cdot \partial E/\partial {\mathbf M} 
}
=
\overline{ p_S }
-
\overline{ p_\alpha }
\, , 
\label{Lanene}
\end{eqnarray}
%----------------------------------------------------------
where 
$p_\alpha
=
\gamma M {\mathcal V}
({\mathbf m} \times {\mathbf H}_{\rm eff})
\cdot 
(\alpha {\mathbf m} \times {\mathbf H}_{\rm eff} + {\mathbf h})$
is the sum of the power dissipated by the Gilbert damping and that generated by the thermally fluctuating magnetic field. 
The average is 
$\langle \overline{p_\alpha} \rangle 
= G_\alpha (\hbar \Omega)^2 \sin^2 \theta$, 
where 
$G_\alpha = \pi \alpha M {\mathcal V}/(h \mu_{\rm B})$. 
The variance is proportional to the temperature times the average 
$\langle \delta \overline{p_\alpha(t)} \, \delta \overline{p_\alpha(t')} \rangle = 2 k_{\rm B} T \langle \overline{p_\alpha} \rangle \delta (t-t')$ 
($\delta \overline{p_\alpha} \equiv \overline{p_\alpha} - \langle \overline{p_\alpha} \rangle$), 
which is a consequence of the fluctuation-dissipation theorem~\cite{Brown}. 

The power gain by the spin-torque is 
%----------------------------------------------------------
\begin{eqnarray}
\overline{
p_S
}
=
2
\mu_B
\overline{
{\mathbf I}_S
\cdot 
{\mathbf H}_{\rm eff}
}
/\hbar
=
\Omega
\overline{
I_{Sz}
}
\, . 
\label{powersptr}
\end{eqnarray}
%----------------------------------------------------------
For the uniaxial anisotropy case, only the $z$ component of the spin torque is necessary 
$\overline{ I_{Sz} }
=
{\mathcal I}
\sin^2 \theta$. 
Our main task is to determine the probability distribution of the fluctuating $\overline{I_{Sz}}$ to be consistent with the FT for a small engine (\ref{ftsmen}). 

{\it Fluctuation theorem for non-conservative force}  -- 
During a time interval $\Delta t$, which is short but sufficiently longer than the period of the precession $2 \pi/\Omega$, $n$ electrons are transmitted through the nano-magnet from left to right leads and the $s$ electron spins flip  from $\uparrow$ to $\downarrow$. 
They are given by 
$n=\int_t^{t+\Delta t}dt' \overline{I(t')}/e$, where $I$ is the charge current, 
and $s=\int_t^{t+\Delta t}dt' \overline{I_{Sz}(t')}/\hbar$. 
When the energy change is slow enough, we can calculate the joint probability distribution $P_{\Delta t}(n,s)$ using the full-counting statistics under the adiabatic pumping with the pumping frequency $\Omega$~\cite{Andreev,Kamenevbook}. 
The scaled cumulant generating function (SCGF) ${\mathcal F}_G$ is introduced as 
%----------------------------------------------------------
\begin{align}
\sum_{n,s}
P_{\Delta t}(n,s; \Omega)
e^{i \lambda n + i \chi s}
&\approx
e^{\Delta t {\mathcal F}_G(\lambda,\chi; \Omega)}
\, , 
\label{foutra1}
\end{align}
%----------------------------------------------------------
where $\lambda$ and $\chi$ are counting fields for the numbers of transmitted electrons and flipped spins. 
Electrons in the left ferromagnetic lead and those in the right metal lead obey the Fermi distribution:
%----------------------------------------------------------
$f_r(E) = 1/[e^{\beta (E-\mu_r)} +1]$ 
($r=L,R$). 
%----------------------------------------------------------
In equilibrium, the chemical potentials are at the Fermi level $\mu_L=\mu_R=E_{\rm F}$. 
The source drain bias voltage $V$ shifts the chemical potential of the left lead as $\mu_L=E_{\rm F}+e V$. 

For now, to keep the discussion simple and specific, we keep the general form of the SCGF under the adiabatic pumping later, Eq.~(\ref{eqn:cgfsmat}), and assume that the nano-magnet is ferromagnetic-insulating, although the current experiments use an insulator/metallic ferromagnet nano-structure~\cite{Suzuki}. 
The SCGF acquires the bi-directional Poisson form~\cite{supl1}: 
%----------------------------------------------------------
\begin{align}
{\mathcal F}_G(\lambda,\chi; \Omega)
&=
\sum_{\nu,\nu' = \pm}
\Gamma_{\nu \nu'}(\Omega)
(e^{i \nu \lambda + i \nu' \chi}-1)
\nonumber \\
&
+
\sum_{\pm}
\Gamma_{\pm}
(e^{\pm i \lambda}-1)
\, .
\label{cgfbidpoi}
\end{align}
%----------------------------------------------------------
The first line corresponds to the spin-flip tunneling process. 
The tunneling rate is 
%----------------------------------------------------------
\begin{align}
\Gamma_{\nu \nu'}(\Omega)
&=
\sin^2 \theta 
\, 
G_{\nu \nu'} 
\, 
\frac{\nu eV - \nu' \hbar \Omega}
{1-e^{-\beta (\nu eV - \nu' \hbar \Omega)}}
\, , 
\end{align}
%----------------------------------------------------------
where 
$G_{++}=G_{--}=G_+$
and 
$G_{+-}=G_{-+}=G_-$
are spin-flip tunnel conductances. 
Their dimension is $h^{-1}$ and 
$G_{+/-}$ connects 
$L \uparrow/L \downarrow$ 
and 
$R \downarrow/R \uparrow$ states. 
The second line of Eq.~(\ref{cgfbidpoi}) corresponds to the spin-preserving tunneling process. 
%----------------------------------------------------------
\begin{align}
\Gamma_\nu
=&
[
G_{\rm P} 
\cos^2 (\theta/2)
+
G_{\rm AP}
\sin^2 (\theta/2)
\nonumber \\
&
-
\sin^2 \theta
(
G_+ + G_-
)
]
\, 
(\nu eV)/(1-e^{-\nu \beta eV})
\, . 
\label{cgfsp}
\end{align}
%----------------------------------------------------------
Similar to the free energy~\cite{Bruno}, from the derivative of the SCGF, we can calculate the charge/spin current. 
For example, we obtain the spin-valve expression~\cite{Slonczewski1}, 
%----------------------------------------------------------
\begin{align}
\frac{\langle \overline{I} \rangle}{e}
=
\left.
\frac{\partial {\mathcal F}_G(\lambda,0;0)}
{
\partial (i \lambda)
}
\right|_{\lambda=0}
=
\left(
G_{\rm P} 
\cos^2 \frac{\theta}{2}
+
G_{\rm AP}
\sin^2 \frac{\theta}{2}
\right) e V
\, ,
\nonumber
\end{align}
%----------------------------------------------------------
where $G_{\rm P}$ and $G_{\rm AP}$ are conductances in parallel and anti-parallel alignments. 

The SCGF is symmetric under the time reversal in the backward driving $\Omega \to -\Omega$. 
It leads the spintronic FT~\cite{UI,Lopez}:
%----------------------------------------------------------
\begin{align}
{\mathcal F}_G(\lambda,\chi; \Omega)
=
{\mathcal F}_{G , R}
(-\lambda + i \beta eV, 
\chi + i \beta \hbar \Omega;
-\Omega)
\, ,
\label{spft}
\end{align}
%----------------------------------------------------------
where the subscript $R$ means that the magnetizations are also reversed, ${\mathbf M} \to -{\mathbf M}$ and ${\mathbf M}_L \to -{\mathbf M}_L$ (which results in $G_+ \leftrightarrow G_-$). 
After the inverse Fourier transform and identifying the work as $w=s \hbar \Omega$ [see Eq.~(\ref{powersptr})] and the Joule heat as $q=n eV$, we obtain the FT for a small engine (\ref{ftsmen}). 
Our SCGF (\ref{cgfbidpoi}) together with the Langevin equation in the energy coordinate (\ref{Lanene}) enables us to calculate the switching exponent consistent with the FT.

\begin{figure}[ht]
\begin{center}
\includegraphics[width=1 \columnwidth]{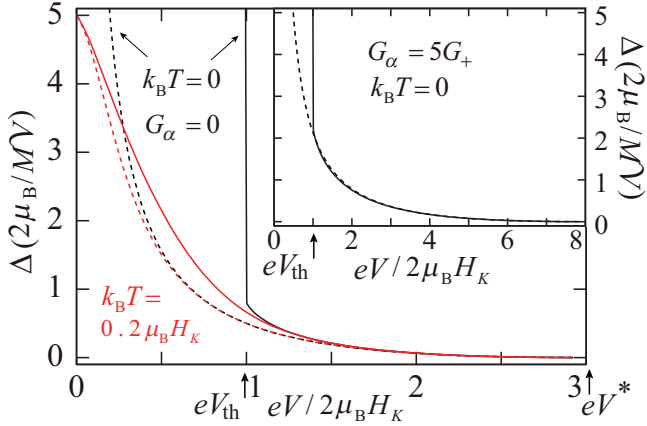}
\caption{
The bias-voltage dependence of the switching exponent  $\Delta$ for $\alpha=0$ and 
$G_-= 2 G_+$. 
The two solid lines are for different temperatures $k_{\rm B}T/(2 \mu_{\rm B} H_K)=0$ and 0.1. 
The inset shows a plot for finite $G_\alpha=5 G_+$. 
The dashed lines indicate an Arrhenius-like law, Eq.~(\ref{appexp}). 
The critical voltages are $eV^*/(2 \mu_{\rm B} H_K)=3$ and 8 for $G_\alpha=0$ and $5 G_+$. 
}
\label{sw}
\end{center}
\end{figure}

{\it Magnetization switching} -- 
The average value of the power (\ref{powersptr}) is given by 
%----------------------------------------------------------
\begin{align}
\langle 
\overline{p_S} 
(\Omega)
\rangle
=
\left.
\hbar 
\Omega
\frac{\partial {\mathcal F}_G(0,\chi;\Omega)}
{
\partial (i \chi)
}
\right|_{\chi=0}
=
\Omega
I_{Sz}^{\Omega=0}
-
p_{\rm pump}
\, . 
\label{blaeqn}
\end{align}
%----------------------------------------------------------
The first term is the power gain by the spin torque: 
$I_{Sz}^{\Omega=0}
=\hbar \sin^2 \theta \, 
(G_+ - G_-)
eV$. 
The second term is the power dissipation by the adiabatic pumping of spin angular momentum~\cite{Brataas}: 
$p_{\rm pump}
=\sin^2 \theta \, 
(\hbar \Omega)^2
(G_+ + G_-)$, 
which accounts for the backaction effect. 
We assume that initially the magnetizations are in antiparallel alignment, $m_z=\cos \theta=-1$. 
Then for $G_+ < G_-$, which means that the spin-flip process $L \downarrow \to R \uparrow$ is the majority process, at positive $eV$, there exists a frequency $\Omega^*$ at which the power gain and the power dissipation balance: 
$
\langle 
\overline{p_S} 
(\Omega^*)
\rangle
=
\langle 
\overline{p_\alpha}
(\Omega^*)
\rangle
$. 
The condition leads,
$
\hbar \Omega^*
=
(G_+ - G_-) eV
/(G_+ + G_- + G_\alpha)
$. 
When the magnitude of the precession frequency at $m_z=-1$, $-\Omega=\gamma H_K$ becomes smaller than $-\Omega^*$, $m_z$ starts to increase to $m_z=0$ and eventually reaches $m_z=1$. 
The critical voltage $eV^*$ above which the magnetization is reversed even in the absence of thermal fluctuations and spin-torque shot noise is 
%----------------------------------------------------------
\begin{eqnarray}
\frac{eV^*}
{2 \, \mu_{\rm B} H_K}
=
\frac{G_+ + G_+ + G_\alpha}{G_- - G_+}
\, . 
\end{eqnarray}
%----------------------------------------------------------
Since the spin-torque shot noise is intrinsic and remains even at zero temperature, the nano-magnet switches probabilistically under $eV^*$. 
A convenient way to calculate such switching probability is the path-integral approach of the Langevin equation (\ref{Lanene})~\cite{sup2}. 
The switching probability $P_\tau$ is the conditional probability to find 
$m_z=-1$ ($E=-M H_K {\mathcal V}/2$) at $t=0$ and $m_z=0$ ($E=0$) at $t=\tau$. 
It is given by
%----------------------------------------------------------
\begin{align}
P_\tau
=&
\int {\mathcal D} \xi
\int_{E(0)=-M H_K {\mathcal V}/2}^{E(\tau)=0} 
{\mathcal D} E
\, e^{i {\mathcal S} }
\, , 
\nonumber 
\\
i {\mathcal S}
=&
-
\int_0^\tau dt
\left[
i \xi(t) \dot{E}(t) 
- 
{\mathcal F}_G
(0,\xi(t) \hbar \Omega(t);\Omega(t))
\right.
\nonumber \\
&
\left.
- 
{\mathcal F}_\alpha
(-\xi(t))
\right]
\, ,
\label{effectiveS}
\end{align}
%----------------------------------------------------------
where we added the SCGF of Gaussian thermal noise, 
%----------------------------------------------------------
\begin{align}
{\mathcal F}_\alpha(\xi)
=
G_\alpha
\sin^2 \theta 
(\hbar \Omega)^2
i \xi (1+i \xi/\beta)
\, . 
\nonumber
\end{align}
%----------------------------------------------------------
Since the number of magnetic moments in the nano-magnet $M {\mathcal V}/\mu_{\rm B}$ is typically large, we utilize the optimal-path approximation. 
The resulting switching probability acquires the form of Eq.~(\ref{swipro}) with the switching exponent: 
%----------------------------------------------------------
\begin{eqnarray}
\Delta
=
-
i {\mathcal S}^*
=
-
\frac{M {\mathcal V} }{2 \mu_{\rm B}}
\int_{-\gamma H_K}^{\Omega^*}
d \Omega 
\frac{i \chi^*}
{\gamma H_K}
\, . 
\label{expopt}
\end{eqnarray}
%----------------------------------------------------------
When the Gilbert damping is absent $\alpha=0$, 
$i \chi^* = \ln [(\Gamma_{+-}+\Gamma_{--})/(\Gamma_{++}+\Gamma_{-+})]$. 
The solid lines in Fig.~\ref{sw} are the switching exponents as a function of the bias voltage at a finite temperature and at zero temperature. 
We find that, at zero temperature below $eV_{\rm th}=2 \mu_{\rm B} H_K$, the exponent diverges, which means that the switching is completely blocked. 
This is because the spin flip process $\downarrow \to \uparrow$ is blocked: 
$\Gamma_{-+}+\Gamma_{--}=0$. 
At finite temperature, this divergence disappears and at $eV=0$, we obtain the Arrhenius law:
$\Delta = M H_K {\mathcal V}/(2 k_{\rm B} T)$. 
The inset shows results at a finite $\alpha$. 
We see that the divergence remains.

Close to the critical voltage, we approximate 
$
i \chi^*
\approx
(\hbar \Omega - \hbar \Omega^*)/(k_{\rm B} \, T_{\rm eff})
$
and obtain the Arrhenius-like form 
%----------------------------------------------------------
\begin{eqnarray}
\Delta 
=
\frac{M H_K{\mathcal V}}{2 k_{\rm B} T_{\rm eff}}
\left(
\frac{V^*-V}{V^*}
\right)^2
\, , 
\label{appexp}
\end{eqnarray}
%----------------------------------------------------------
which quadratically depends  on the distance from the critical voltage. 
The effective temperature, 
%----------------------------------------------------------
\begin{align}
T_{\rm eff}
=
\frac{
\sum_\pm
G_\mp (eV \pm \hbar \Omega^*) \coth \frac{eV \pm \hbar \Omega^*}{2 k_{\rm B} T}
+
2 k_{\rm B} T \, 
G_\alpha
}
{2 k_{\rm B} \, (G_+ + G_- + G_\alpha)}
\, ,  
\nonumber
\end{align}
%----------------------------------------------------------
is reduced to the real temperature  $T_{\rm eff} \approx T$ for high temperatures, $eV,eV_{\rm th} \ll k_{\rm B} T$. 
Then Eq.~(\ref{appexp}) reproduces the previous result~\cite{TI,TUMGI}. 
At zero temperature and $G_\alpha=0$, $T_{\rm eff} \approx 2 G_+ G_- eV/(G_+ + G_-)^2$, which is proportional to the bias voltage~\cite{Nunez}, indicating that the spin-torque shot noise is the dominant source of fluctuations around the critical voltage.  
The dashed lines in Fig.~\ref{sw} show Eq.~(\ref{appexp}). 
They fit well for finite temperature or around the critical voltage. 

When the volume becomes very small, i.e., ${\mathcal V} \sim \mu_B/M$, we have to go beyond the optimal path approximation~\cite{Kanazawa,Kanazawa1}. 
In such cases , the time scales of the source of Gaussian noise and that of Poisson noise should be treated carefully~\cite{Kanazawa1}. 

%----------------------------------------------------------
%----------------------------------------------------------

{\it Full-counting statistics under the adiabatic pumping} -- 
An electron transferred through the nano-magnet is affected by its precession motion. 
This scattering process is described by a time-dependent $S$-matrix: 
%----------------------------------------------------------
\begin{align}
{\mathbf S}
(\theta,\phi(t))
&=
e^{-i \phi(t) \sigma_z/2} 
{\mathbf S}(\theta)
e^{i \phi(t) \sigma_z/2} 
\\
{\mathbf S}(\theta)
&=
\left(
\begin{array}{cc}
{\mathbf r}(\theta)  & {\mathbf t}'(\theta) \\
{\mathbf t}(\theta)  & {\mathbf r}'(\theta) 
\end{array}
\right)
\, , 
\end{align}
%----------------------------------------------------------
where the Pauli matrix $\sigma_z$ acts in the spin space and $\phi(t) \approx \Omega \, t + \phi(0)$. 
${\mathbf r}$ and ${\mathbf t}$ (${\mathbf r}'$ and ${\mathbf t}'$) are $2 \times 2$ matrix of the spin-dependent reflection amplitudes and that of the spin-dependent transmission amplitudes for an incoming wave from the left (right) lead. 
For example, the element $t_{\sigma' \sigma}$ describes an electron transmission from the spin $\sigma$ state in the left lead to the spin $\sigma'$ state in the right lead. 

The SCGF~(\ref{foutra1}) is expressed by using the $S$-matrix as~\cite{Andreev,Kamenevbook}, 
%----------------------------------------------------------
\begin{align}
{\mathcal F}_G&(\lambda,\chi;\Omega)
=
\sum_\ell
\int 
\frac{d E}{h}
\ln \det
\Big [
{\mathbf 1}-
{\mathbf f}(E) 
\Big ( 
{\mathbf 1} 
\nonumber \\
& -
e^{i \mat{\lambda} + i \chi \sigma_z/2}
{\mathbf S}^\ell(\theta;E)^\dagger 
e^{-i \mat{\lambda}-i \chi \sigma_z/2}
{\mathbf S}^\ell(\theta;E)
\Big )
\Big ]
, 
\label{eqn:cgfsmat}
\end{align}
%----------------------------------------------------------
where ${\mathbf S}^\ell(E)$ is the $S$-matrix for the $\ell$-th transverse channel. 
The counting field matrix $\mat{\lambda}={\rm diag}(\lambda,\lambda,0,0)$ counts the number of electrons flowing out of the left lead. 
The precession motion effectively splits the $\uparrow$-spin and $\downarrow$-spin chemical potentials of the 2 leads after the gauge transform, 
${\bf f}(E)={\rm diag}(f_L(E+\hbar \Omega/2),f_L(E-\hbar \Omega/2),f_R(E+\hbar \Omega/2),f_R(E-\hbar \Omega/2))$. 
The spin-splitting of the chemical potentials is a result of the backaction, which is crucial to be consistent with the FT~\cite{UGMSK}. 
It also blocks the spin-flip tunneling process $\downarrow \to \uparrow$ under the threshold voltage. 

Although we considered a simple model, it is also possible to calculate the $S$-matrix using a realistic model. 
Then, from Eq.~(\ref{eqn:cgfsmat}), we obtain Eq.~(\ref{blaeqn}) expressed with general $I_{Sz}^{\Omega=0}$ and $p_{\rm pump}$, 
%----------------------------------------------------------
\begin{align}
I_{Sz}^{\Omega=0}
=&
\frac{eV}{4 \pi}
\sum_{\ell , \sigma=\uparrow,\downarrow}
(
|r^\ell_{\downarrow \, \sigma}(\theta,\phi;E_F)|^2
+
|t^\ell_{\downarrow \, \sigma}(\theta,\phi;E_F)|^2
\nonumber \\
&-
|r^\ell_{\uparrow \, \sigma}(\theta,\phi;E_F)|^2
-
|t^\ell_{\uparrow \, \sigma}(\theta,\phi;E_F)|^2
)
\, , 
\nonumber 
\\
p_{\rm pump}
=&
\frac{\hbar \, \Omega^2}{4 \pi}
\,
\sum_{\ell}
{\rm tr}
(
\partial_\phi {\mathbf S}^{\ell}(\theta,\phi;E_F)
 \, 
\partial_\phi {\mathbf S}^{\ell}(\theta,\phi;E_F)^\dagger
)
\, , 
\nonumber 
\end{align}
%----------------------------------------------------------
in the leading order of $eV$ and $\Omega$. 
It is straightforward to take the channel mixing scattering into account. 
Our $p_{\rm pump}$ reproduces Ref.~\onlinecite{Brataas}. 

{\it Summary} -- 
We demonstrate the switching probability driven by fluctuating non-conservative spin-torque. 
The theory of the full-counting statistics under the adiabatic pumping enables us to account for the backaction effect and to obtain a distribution of the fluctuating spin-torque consistent with the fluctuation theorem for a small engine. 
We find the threshold voltage $eV_{\rm th}=2 \mu_{\rm B} H_K$, above which the spin-torque shot noise causes the probabilistic switching. 
Under the threshold the spin-flip tunneling process is blocked because of the backaction and thus the probabilistic switching is suppressed. 

This work was supported by KAKENHI (grant numbers 26400390 and 26220711).

%\end{document}

%\clearpage

\begin{widetext}

%\documentclass[amsmath,amssymb,prl]{revtex4}
%\usepackage[dvips]{graphicx}
%\usepackage{bm}% bold math
%\begin{document}
%\newcommand{\mat}[1]{\mbox{\boldmath$#1$}}

\section{Supplemental Material}

Technical details of derivations of a scattering matrix, a scaled cumulant generating function and a switching exponent. 

\subsection{Scattering matrix and the scaled cumulant generating function}

We derive the $S$-matrix of the ferromagnet/ferromagnetic insulator/normal metal structure. 
We take the $z$-axis perpendicular to the interface and assume translational invariance in the $x$ and $y$ directions. 
The Schr\"odinger equation is
%----------------------------------------------------------
\begin{align}
\left(
-
\frac{\hbar^2}{2m} 
\nabla^2
+
U (z)
\right)
\, \psi (x,y,z)
=
E
\, \psi (x,y,z)
\, , 
\;\;\;\;
U(z)
=
\left \{
\begin{array}{cc}
\mu_{\rm B} {H_{\rm m}}_L \, {\sigma}_z/2 & (z<0) \\
U_0 +  \mu_{\rm B} H_{\rm m} \, {\mathbf m} \cdot \vec{\sigma} & 
(0 \leq z < d) \\
0 & (d \leq z)
\end{array}
\right.
\, , 
\end{align}
%----------------------------------------------------------
where 
$\vec{\sigma}=({\sigma}_x,{\sigma}_y,{\sigma}_z)$
is the Pauli matrix vector. 
The thickness of the ferromagnetic insulator is $d$ and $U_0>0$ is the potential barrier height. 
The molecular (exchange) fields in the ferromagnetic lead and in the ferromagnetic insulator are ${H_{\rm m}}_L$ and $H_{\rm m}$, respectively. 
The wave function is written as
$\psi(x,y,z)
=
2
\sin (\pi \ell_x x/L)
\sin (\pi \ell_y y/L) \, 
\psi(z)/L$
where the contact area is $0 \leq x,y \leq L$ 
($\ell_y$ and $\ell_z$ are non-negative integers). 
In the $z$ direction, the Schr\"odinger equation reads 
%----------------------------------------------------------
\begin{eqnarray}
\left(
-
\frac{\hbar^2}{2m} 
\frac{\partial^2}{\partial z^2} 
+
U (z)
\right)
\, \psi_{\ell} (z)
=
E_\ell
\, \psi_{\ell} (z)
\, , 
\;\;\;\;
E_\ell
=
E-
\frac{\hbar^2 \pi^2}{2m L^2}
(\ell_y^2 + \ell_z^2)
\, , 
\end{eqnarray}
%----------------------------------------------------------
where we introduced the channel index $\ell=(\ell_x,\ell_y)$. 
The wave number of an electron with the energy $E$ is 
$k_\sigma =\sqrt{2m (E_\ell-\sigma \mu_{\rm B} {H_{\rm m}}_L)}/\hbar$ in the ferromagnetic lead ($z<0$), $i \kappa_\sigma = \sqrt{2m (E_\ell-U_0 - \sigma \mu_{\rm B} H_{\rm m}/2)}/\hbar$ in the ferromagnetic insulator ($0<z<d$)
and $k =\sqrt{2m E_\ell}/\hbar$ in the normal metal lead ($d<z$). 
The $S$-matrix in the leading order of $e^{-\kappa_\sigma d}$ is calculated as 
%----------------------------------------------------------
\begin{eqnarray}
{\mathbf S}(\theta)
=
\left(
\begin{array}{cc}
{\mathbf P} & {\mathbf 0} \\
{\mathbf 0} & {\mathbf P}'
\end{array}
\right)
\left(
\begin{array}{cc}
-{\mathbf 1} - (i {\mathbf A}+ \mat{\tau}^\dagger \mat{\tau}/2) & \mat{\tau}^\dagger \\
\mat{\tau} & {\mathbf 1} + (i {\mathbf A} + \mat{\tau} \mat{\tau}^\dagger/2)
\end{array}
\right)
\left(
\begin{array}{cc}
{\mathbf P} & {\mathbf 0} \\
{\mathbf 0} & {\mathbf P}'
\end{array}
\right)
\, . 
\label{Sm}
\end{eqnarray}
%----------------------------------------------------------
where an Hermite matrix ${\mathbf A}^\dagger = {\mathbf A}$ is not relevant for our model. 
Further, we neglect $H_{\rm m}$ except when it appears in the exponent of $e^{-\kappa_\sigma d}$. 
Then we obtain the following $2 \times 2$ matrix of the spin-dependent transmission amplitude: 
%----------------------------------------------------------
\begin{align}
\mat{\tau}
=
\frac{1}{2}
\left(
\begin{array}{cc}
\tau_{\uparrow \uparrow} + \tau_{\downarrow \uparrow}
+
(\tau_{\uparrow \uparrow}- \tau_{\downarrow \uparrow})
\cos \theta 
&
(\tau_{\uparrow \downarrow} - \tau_{\downarrow \downarrow}) \sin \theta 
\\
(\tau_{\uparrow \uparrow}- \tau_{\downarrow \uparrow}) \sin \theta 
&
\tau_{\uparrow \downarrow} + \tau_{\downarrow \downarrow} +
(\tau_{\uparrow \downarrow} - \tau_{\downarrow \downarrow}) \cos \theta 
\end{array}
\right)
\, ,
\;\;\;\;
\tau_{\sigma \sigma'}
=
4 e^{-\kappa_\sigma d}
\sqrt{
\frac{\kappa_0 k}{\kappa_0^2 +k^2}
\frac{\kappa_0 k_{\sigma'}}{\kappa_0^2 +k_{\sigma'}^2}
}
. 
\end{align}
%----------------------------------------------------------
$2 \times 2$ sub-matrices ${\mathbf P}$, ${\mathbf P}'$ become diagonal and $(\sigma,\sigma)$ component of ${\mathbf P}^2$ and ${{\mathbf P}'}^2$ are $(\kappa_0+i k_\sigma)/(\kappa_0-i k_\sigma)$ and $-i (\kappa_0+i k)/(\kappa_0-i k)$,
where $\kappa_0 =\sqrt{2m (U_0 - E_\ell)}/\hbar$. 

We insert the $S$-matrix (\ref{Sm}) into Eq.~(20) in the main text:
%(\ref{eqn:cgfsmat}), 
%----------------------------------------------------------
\begin{align}
{\mathcal F}_G(\lambda,\chi;\hbar \Omega)
&=
\rho_\parallel
\int 
d E_\parallel
\int 
\frac{d E}{h}
\ln \det
\Big [
{\mathbf 1}+
{\mathbf f}(E) 
\Big ( 
e^{i \mat{\lambda} + i \chi \sigma_z/2}
{\mathbf S}(E-E_\parallel,\theta)^\dagger 
e^{-i \mat{\lambda} - i \chi \sigma_z/2}
{\mathbf S}(E-E_\parallel,\theta)
 - {\mathbf 1} 
\Big )
\Big ]
\, , 
\label{nqcgfsma}
\end{align}
%----------------------------------------------------------
where $\rho_\parallel=2 \pi mL^2/h^2$ is the DOS of the transverse channel. 
Since the energy dependence of $\tau_{\sigma \sigma'}$ is small around the Fermi energy $E_{\rm F}$, it is possible to approximate 
$\tau_{\sigma \sigma'}(E-E_\parallel)
\approx 
\tau_{\sigma \sigma'}(E_F) \exp(-E_\parallel/(2 \delta))$, 
where 
$\delta^{-1} = 2d \, \partial \kappa_0(E_\ell=E_F)/\partial E_\ell$. 
After performing the integral and expanding up to the leading order in $e^{-\kappa_\sigma d}$, we obtain Eq.~(9) in the main text. 
%(\ref{cgfbidpoi}). 
The conductances are
%----------------------------------------------------------
\begin{align}
G_+
&=
\frac{1}{h}
\, 
\rho_\parallel \delta
\left|
\tau_{\downarrow \downarrow}(E_F) - \tau_{\uparrow \downarrow}(E_F)
\right|^2
\, ,
\\
G_-
&=
\frac{1}{h}
\, 
\rho_\parallel \delta
\left|
\tau_{\uparrow \uparrow}(E_F) - \tau_{\downarrow \uparrow}(E_F)
\right|^2
\, ,
\\
G_{\rm P}
&=
\frac{1}{h}
\, 
\rho_\parallel \delta
(
|\tau_{\uparrow \uparrow}(E_F)|^2
+
|\tau_{\downarrow \downarrow}(E_F)|^2
)
\, ,
\\
G_{\rm AP}
&=
\frac{1}{h}
\, 
\rho_\parallel \delta
(
|\tau_{\uparrow \downarrow}(E_F)|^2
+
|\tau_{\downarrow \uparrow}(E_F)|^2
)
\, . 
\end{align}
%----------------------------------------------------------
The reversal of the magnetic moments 
${\mathbf M} \to -{\mathbf M}$
and 
${\mathbf M}_L \to -{\mathbf M}_L$, 
which corresponds to 
$H_{\rm m} \to -H_{\rm m}$ 
and 
${H_{\rm m}}_L \to -{H_{\rm m}}_L$, 
changes the tunneling amplitude to 
$\tau_{\sigma \sigma'} \to \tau_{\sigma' \sigma}$
and thus the conductances to $G_+ \leftrightarrow G_-$.

\subsection{Switching exponent}

We analyze the Langevin equation (6) in the main text 
%(\ref{Lanene}) 
by exploiting the Martin-Siggia-Rose approach (see Section 4 in Ref.~\onlinecite{s_Kamenevbook}). 
We first discretize time $\tau$ into $N=\tau/\Delta t$ steps. 
For now, we neglect the equilibrium power dissipation $p_\alpha$. 
The variation of the energy during a short time step from $t_j=\Delta t \, j$ to $t_{j+1}$ is 
%----------------------------------------------------------
\begin{eqnarray}
E_{j+1} - E_j
 \approx 
\hbar \Omega_j 
s_{j}
\, , 
\;\;\;\;
s_j
=
\int_{t_j }^{t_{j+1}}
dt \, 
\overline{
I_{Sz}(t)
}
\, , 
\end{eqnarray}
%----------------------------------------------------------
where $E_j=E(t_j)$ and 
$\Omega_{j}=\Omega((E_{j+1}+E_j)/2)$. 
The stochastic variable $s_j$ is distributed according to the joint probability distribution Eq.~(8) described in the main text. 
%(\ref{foutra1}). 
The conditional joint probability to find $E_j$ at time $t_j$ and $E_{j+1}$ at $t_{j+1}$ accompanied by $n_j$ electron transmission is given by 
%----------------------------------------------------------
\begin{align}
P_{\Delta t}
(n_j, E_{j+1}|E_j)
&=
\int 
d \epsilon_{S j}
\, 
\delta
(
E_{j+1}
-E_j
-
\epsilon_{S j}
)
\, 
\sum_{s,n}
P_{\Delta t}(n,s; \Omega_j) \, 
\delta 
(\epsilon_{S j} - \hbar \Omega_j s)
\, 
\delta_{n_j,n}
\\
&=
\int_{-\pi}^\pi \frac{d \lambda_j}{2 \pi}
\int \frac{d \xi_j}{2 \pi}
e^{
- i \lambda_j n_j
- i \xi_j 
(
E_{j+1}
-E_j
)
+
{\mathcal F}_G(\lambda_j,\hbar \Omega_j \xi_j;\Omega_j) \Delta t
}
\, . 
\end{align}
%----------------------------------------------------------
Then the conditional joint probability to find $E(0)$ at $t=0$ and $E(\tau)$ at $\tau$ accompanied by $n$ electron transmission is calculated by accumulating joint probabilities for short time steps as follows:
%----------------------------------------------------------
\begin{align}
P_\tau(n, E(\tau)|E(0))
=&
\sum_{n_0, \cdots, n_{N-1}}
\int d E_1 \cdots d E_{N-1}
P_{\Delta t}(n_{N-1},E_{N}|E_{N-1})
%\nonumber \\ & \times P_{\Delta t}(n_{N-2},E_{N-1}|E_{N-2})
\cdots
P_{\Delta t}(n_{0},E_{1}|E_{0})
\, 
\delta_{n, \sum_{j=0}^{N-1} n_j}
\nonumber 
\\
=&
\int_{-\pi}^\pi \frac{d \lambda}{2 \pi}
\int \frac{d \xi_0}{2 \pi} \cdots \frac{d \xi_{N-1}}{2 \pi}
\int d E_1 \cdots d E_{N-1}
e^{\sum_{j=0}^{N-1}
[-i \xi_j (E_{j+1}-E_j) 
+ 
{\mathcal F}_G
(\lambda, \xi_j \hbar \Omega_j;\Omega_j)
\Delta t]
-
i \lambda n
}
\, . 
\end{align}
%----------------------------------------------------------
We can prove the detailed FT by Jarzynski~\cite{s_Jarzynski} based on the FT (12) in the main text 
%(\ref{spft}) 
along the same line of the proof in Ref.~\onlinecite{s_UEAKT}, 
%----------------------------------------------------------
\begin{align}
P_\tau(n,E(\tau)|E(0))/P_{R,\tau}(-n,E(0)|E(\tau))
=
e^{\beta [n eV-E(\tau)+E(0)]}
\, ,
\end{align}
%----------------------------------------------------------
In order to account for the equilibrium power dissipation, we can just replace ${\mathcal F}_G$ with 
${\mathcal F}_G(\lambda, \xi \hbar \Omega;\Omega)+{\mathcal F}_\alpha(-\xi)$. 
Further, for calculating the switching rate, we can sum over $n$, 
$P_\tau(E(\tau)|E(0)) \equiv \sum_n P_\tau(n,E(\tau)|E(0))$. 
Then in the continuous limit, $\Delta t \to 0$, we obtain the path-integral form Eq.~(15) in the main text. 
%(\ref{effectiveS}). 

Since $E, G_\alpha \propto {\mathcal V}=L^2d$ and ${\mathcal F}_G \propto L^2/d$, for a modestly large nano-magnet, it is possible to perform the optimal path approximation~\cite{s_Tobiska1,s_Sukhorukov,s_Billings,s_Kamenevbook}. 
Namely, from the variational principle, we derive the ``canonical equation of motion": 
%----------------------------------------------------------
\begin{eqnarray}
\dot{E}
=
\frac{\partial {\mathcal F}}{\partial (i \xi)}
\, , 
\;\;\;\;
i \dot{\xi}
=
-
\frac{\partial {\mathcal F}}{\partial E}
\, . 
\label{hameq}
\end{eqnarray}
%----------------------------------------------------------
The ``momenta" $i \xi$ measures the strength of the fluctuations. 
$i \xi=0$ corresponds to the noiseless case, which is always an optimal path. 
The equation of motion possesses the integral of motion, which is the ``energy," ${\mathcal F}$. 
Since the normalization condition ensures ${\mathcal F}(\xi=0;\Omega)=0$, the optimal paths always satisfy ${\mathcal F}(\xi;\Omega)=0$. 

We are interested in an optimal path that starts from $(E,i \xi)=(-M H_K {\mathcal V}/2,0)$ and reaches $(E,i \xi)=(0,0)$.  
For $\alpha=0$, we find 4 simple solutions satisfying ${\mathcal F}(\xi;\Omega)=0$: 
%----------------------------------------------------------
\begin{eqnarray}
\hbar \Omega(E)=\pm 2 \mu_{\rm B} H_K \, , 
\;\;\;\;
i \hbar \Omega(E) \xi^*
= 
\ln \frac{\Gamma_{+-}+\Gamma_{--}}{\Gamma_{++}+\Gamma_{-+}} \, , 
\;\;\;\;
i \xi^*=0
\, . 
\end{eqnarray}
%----------------------------------------------------------
Figure \ref{separatrix} (a) shows the optimal paths. 
The horizontal axis is $\Omega=-\sqrt{-2 \gamma^2 H_K E/(M {\mathcal V})}$
and thus $E=-M H_K {\mathcal V}/2$ and $E=0$ correspond to $\hbar \Omega=-2 \mu_{\rm B}H_K$ and $\hbar \Omega=0$, respectively. 
Arrows indicate the directions of motion determined from Eq.~(\ref{hameq}). 
The initial state is at M, i.e., 
$(\hbar \Omega, i \xi \hbar \Omega)=(-2 \mu_{\rm B} H_K,0)$, 
and the final state is at T, i.e.,
$(\hbar \Omega, i \xi \hbar \Omega)=(0,0)$. 
The optimal path is $ {\rm M} \to {\rm M}' \to {\rm U} \to {\rm T}$, 
where the intermediate state U is 
$(\hbar \Omega, i \xi \hbar \Omega)=(\hbar \Omega^*, 0)$. 
The action along this path is calculated as
%----------------------------------------------------------
\begin{eqnarray}
i {\mathcal S}^*
=
-
\int_{-M H_K {\mathcal V}/2}^{E(\Omega^*)}
\!\!\!\!
dE \, (i \xi^*)
=
\frac{M {\mathcal V}}{2 \mu_{\rm B} \gamma H_K}
\int_{-2 \mu_{\rm B} H_K}^{\Omega^*}
\!\!\!\!
d \Omega \, i  \xi^* \hbar \Omega
\equiv
-
\Delta 
\, . 
\label{eqn:integ}
\end{eqnarray}
%----------------------------------------------------------
The integral $\int \!\! d \Omega \, i  \xi^* \hbar \Omega$ gives the area of the shaded region in Fig. \ref{separatrix} (a). 
This equation leads to Eq.~(16) in the main text 
%(\ref{expopt}) 
and the switching probability up to the single instanton contribution, 
$P_\tau(E(\tau)=0|E(0)=-M H_K {\mathcal V}/2) \approx e^{-\Delta}$. 

At zero temperature, the integral (\ref{eqn:integ}) can be performed easily. 
For $eV_{\rm th}=2 \mu_{\rm B} H_K < eV < eV^*$, we obtain
%----------------------------------------------------------
\begin{align}
i {\mathcal S}^*
=
\frac{M {\mathcal V}}{2 \mu_{\rm B}}
\left \{ 
\ln \!
\frac{ G_- (eV - 2 \mu_{\rm B} H_K)}{G_+ (eV+2 \mu_{\rm B} H_K)}
+
\frac{eV}{2 \mu_{\rm B} H_K}
\ln \!
\frac{4 G_+ G_- (eV)^2}
{(G_+ + G_-)^2[(eV)^2 - (2 \mu_{\rm B} H_K)^2]}
\right \}
\, . 
\end{align}
%----------------------------------------------------------
For $eV<eV_{\rm th}$, it diverges to $i {\mathcal S}^* = - \infty$, which means that the switching is completely blocked. 
Figure~\ref{separatrix} (b) shows the optimal path at $eV=eV_{\rm th}$. 
M' approaches $(\hbar \Omega, i \xi \hbar \Omega)=(0,-\infty)$ in the limit of zero temperature, and the area of the shaded region diverges. 
For $G_\alpha \neq0$, the optimal path is modified and we determine it numerically. 

With increasing bias voltage, the shaded area decreases and eventually M' and U meet at M [Fig.~\ref{separatrix} (c)]. 
The exponent and the switching probability become $i {\mathcal S}^*=0$ and $P_\tau \approx 1$. 
This critical condition is achieved at 
$\hbar \Omega^*=-2 \mu_{\rm B} H_K$ ,
which is equivalent to the balance condition 
$
\langle p_S 
(\Omega^*)
\rangle
=
\langle p_\alpha
(\Omega^*)
\rangle
$. 
Around the critical point (for $G_\alpha \neq0$), we can expand $\xi^*$ around 
$\Omega=\Omega^*$ and $\xi=0$, up to the lowest order as 
%----------------------------------------------------------
\begin{align}
i \xi^* \hbar \Omega
\approx
\frac{\hbar \Omega - \hbar \Omega^*}{k_{\rm B} T_{\rm eff}}
\, . 
\nonumber
\end{align}
%----------------------------------------------------------
By plugging this expression into Eq.~(\ref{eqn:integ}), we obtain Eq.~(17) in the main text. 
%(\ref{appexp}). 

%----------------------------------------------------------
\begin{figure}[ht]
\begin{center}
\includegraphics[width=1 \columnwidth]{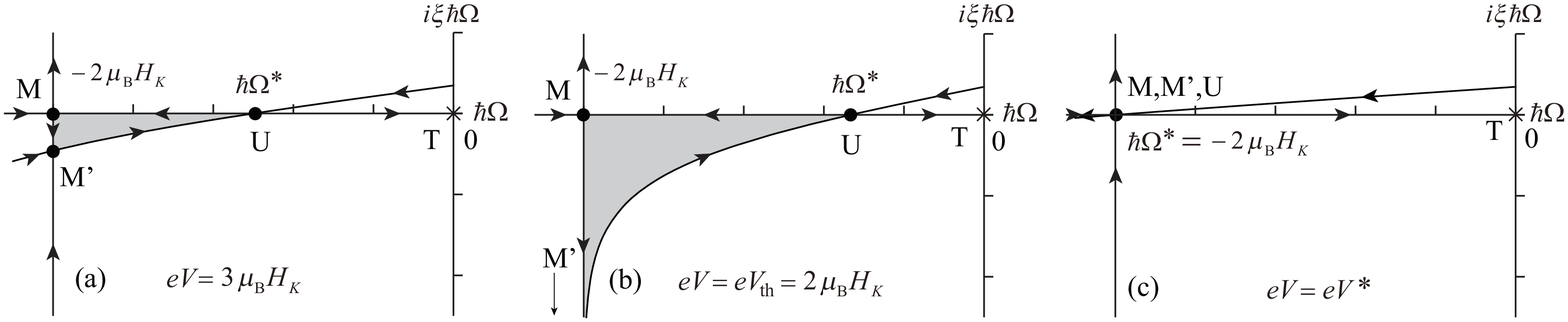}
\caption{
The optimal paths (a) for $eV=3 \mu_{\rm B} H_K$, 
(b) for the threshold voltage $eV=eV_{\rm th}=2 \mu_{\rm B} H_K$
and 
(c) for the critical voltage $eV=eV^*=6 \mu_{\rm B} H_K$.  
The parameters are as follows: 
$G_-=2 G_+$, 
$G_\alpha=0$ 
and 
$k_{\rm B} T=0$. 
}
\label{separatrix}
\end{center}
\end{figure}
%----------------------------------------------------------

\end{widetext}


\begin{thebibliography}{35}

\bibitem{Seifert}
U. Seifert, 
Rep. Prog. Phys. {\bf 75} (2012) 126001. 

\bibitem{EHM} 
M. Esposito, U. Harbola, and S. Mukamel, Rev. Mod. Phys. {\bf 81}, 1665 (2009).
 
\bibitem{CHT} 
M. Campisi, P. H\"{a}nggi, and M. Talkner, Rev. Mod. Phys. {\bf 83}, 771 (2011).

\bibitem{Tobiska}
J. Tobiska and Yu. V. Nazarov, 
Phys. Rev. B {\bf 72}, 235328 (2005). 

\bibitem{FB}
H. F\"{o}rster and M B\"{u}ttiker, Phys. Rev. Lett. {\bf 101}, 136805 (2008).

\bibitem{SU}
K. Saito and Y. Utsumi, 
Phys. Rev. B {\bf 78}, 115429 (2008). 

\bibitem{Andrieux}
D. Andrieux, P. Gaspard, T. Monnai, and S. Tasaki, New J. Phys. {\bf 11}, 
043014 (2009).

\bibitem{UI}
Y. Utsumi, H. Imamura, J. Phys.: Conf. Ser. {\bf 200}, 052030 (2010). 

\bibitem{Lopez}
R. L\'opez, J.-S. Lim, D. S\'anchez, 
Phys. Rev. Lett. 108, 246603 (2012); 
J.-S. Lim, D. S\'anchez, R. L\'opez, 
arXiv:1208.4746. 

\bibitem{Sinitsyn}
N. A. Sinitsyn, 
J. Phys. A: Math. Theor. {\bf 44}, 405001 (2011). 

\bibitem{Campisi}
M. Campisi, arXiv:1403.8040. 

\bibitem{Verley}
G. Verley, T. Willaert, C. Van den Broeck, and M. Esposito, 
arXiv:1404.3095. 

%\bibitem{Kramers}
%H. A. Kramers, Physica {\bf 7}, 284 (1940). 

\bibitem{HTB}
P. H\"anggi, P. Talkner, and M. Borkovec, 
Rev. Mod. Phys. {\bf 62}, 251 (1990). 
 

\bibitem{Slonczewski}
J. C. Slonczewski, 
J. Magn. Magn. Mat. {\bf 159}, L1-L7 (1996). 

\bibitem{Apalkov} 
D. M. Apalkov, P. B. Visscher, 
Phys. Rev. B {\bf 72}, 180405 (2005). 

\bibitem{TI}
T. Taniguchi, H. Imamura, 
Phys. Rev. B {\bf 83}, 054432 (2011). 

\bibitem{Pinna}
D. Pinna, A. D. Kent, D. L. Stein, 
Phys. Rev. B {\bf 88}, 104405 (2013). 

\bibitem{TUMGI}
T. Taniguchi, Y. Utsumi, M. Marthaler, D. S. Golubev, H. Imamura, 
Phys. Rev. B {\bf 87}, 054406 (2013); 
T. Taniguchi, M. Shibata, M. Marthaler, Y. Utsumi, H. Imamura, Appl. Phys. Express {\bf 5} 063009 (2012). 

\bibitem{Suzuki} 
Y. Suzuki, A. A. Tulapurkar, and C. Chappert, 
in {\it Nanomagnetism and Spintronics}, 
edited by T. Shinjo (Elsevier, Amsterdam, 2009). 

\bibitem{Blanter} 
Ya. M. Blanter, M. B\"uttiker, Phys. Rep. {\bf 336}, 1 (2000).

\bibitem{Foros}
J. Foros, A. Brataas, Y. Tserkovnyak, G. E. W. Bauer, 
Phys. Rev. Lett. {\bf 95}, 016601 (2005). 

\bibitem{Nunez}
A. S. N\'u\~nez, R. A. Duine, 
Phys. Rev. B {\bf 77}, 054401 (2008). 

\bibitem{Chudnovskiy}
A. L. Chudnovskiy, J. Swiebodzinski, and A. Kamenev, 
Phys. Rev. Lett {\bf 101}, 066601 (2008). 


\bibitem{Levitov} 
L. S. Levitov, H.-W. Lee, and G. B. Lesovik, 
Journal of Mathematical Physics, {\bf 37}, 4845 (1996).

\bibitem{Andreev}
A. Andreev, A. Kamenev, 
Phys. Rev. Lett. {\bf 85}, 1294 (2000).

\bibitem{Kamenevbook}
A. Kamenev, 
{\it Field Theory of Nonequilibrium Systems} (Cambridge University Press, Cambridge, 2011). 

%\bibitem{Tserkovnyak} Y. Tserkovnyak, A. Brataas, and G. E. W. Bauer, Phys. Rev. Lett. {\bf 88}, 117601 (2002).  

\bibitem{Brataas}
A. Brataas, Y. Tserkovnyak, and G. E. W. Bauer, 
Phys. Rev. Lett. {\bf 101}, 037207 (2008);  
Phys. Rev. B {\bf 84}, 054416 (2011). 

\bibitem{Brown}
W. F. Brown, 
Phys. Rev. {\bf 130}, 1677 (1963). 


\bibitem{supl1}
See Supplemental Material for technical details. 

\bibitem{Bruno}
P. Bruno,
Phys. Rev. B {\bf 52}, 411 (1995).  

\bibitem{Slonczewski1}
J. C. Slonczewski, 
Phys. Rev. B {\bf 39} 6995 (1989). 

\bibitem{sup2}
See Supplemental Material for technical details. 

\bibitem{UGMSK}
Y. Utsumi, D. S. Golubev, M. Marthaler, Gerd Sch\"on, and K. Kobayashi, 
Phys. Rev. B {\bf 86}, 075420 (2012). 

\bibitem{Kanazawa}
K. Kanazawa, T. Sagawa, H. Hayakawa, 
Phys. Rev. Lett. {\bf 108}, 210601 (2012). 

\bibitem{Kanazawa1}
K. Kanazawa, T. G. Sano, T. Sagawa, H. Hayakawa, arXiv:1407.5267. 


\end{thebibliography}

\begin{thebibliography}{6}


\bibitem{s_Kamenevbook}
A. Kamenev, 
{\it Field Theory of Nonequilibrium Systems} (Cambridge University Press, Cambridge, 2011). 

\bibitem{s_Jarzynski}
C. Jarzynski, 
J. Stat. Phys. {\bf 98}, 77 (2000). 

\bibitem{s_UEAKT}
Y. Utsumi, O. Entin-Wohlman, A. Aharony, T. Kubo, and Y. Tokura, 
Phys. Rev. B {\bf 89}, 205314 (2014). 

\bibitem{s_Tobiska1}
J. Tobiska and Yu. V. Nazarov, 
Phys. Rev. Lett. {\bf 93}, 106801 (2004). 

\bibitem{s_Sukhorukov}
E. V. Sukhorukov and A. N. Jordan, 
Phys. Rev. Lett. {\bf 98}, 136803 (2007). 

\bibitem{s_Billings}
L. Billings, I. B. Schwartz, M. McCrary, A. N. Korotkov, and M. I. Dykman, 
Phys. Rev. Lett. {\bf 104}, 140601 (2010). 
\end{thebibliography}
\end{document}